\documentclass[useAMS,usenatbib]{mn2e}
\usepackage{graphicx}
\usepackage{natbib}
\usepackage{amssymb}
\usepackage{setspace}
\usepackage{amssymb}
\usepackage{epsfig,color}
\usepackage{multicol}

\long\def\symbolfootnote[#1]#2{\begingroup%
\def\thefootnote{\fnsymbol{footnote}}\footnote[#1]{#2}\endgroup} 

\newcommand{\sqc}{cm$^{-2}$}                   % cm^-2
\newcommand{\cc}{cm$^{-3}$}                    % cm^-3

\title[Modelling the structure of molecular clouds: I.]{Modelling the structure of molecular clouds: I. A multi-scale energy equipartition}
\author[Veltchev, Donkov \& Klessen]
{	
\parbox{\textwidth}{Todor V. Veltchev$^{1,2}$\thanks{E-mail: \texttt{eirene@phys.uni-sofia.bg}}, Sava Donkov$^3$, and Ralf S. Klessen$^2$}\vspace{0.4cm}\\
 \parbox{\textwidth}{
   $^1$University of Sofia, Faculty of Physics, 5 James Bourchier Blvd., 1164 Sofia, Bulgaria\\
   $^2$Universit\"at Heidelberg, Zentrum f\"ur Astronomie, Institut f\"ur Theoretische Astrophysik, Albert-Ueberle-Str. 2, 69120 Heidelberg, Germany\\
   $^3$Department of Applied Physics, Technical University, 8 Kliment Ohridski Blvd., 1000 Sofia, Bulgaria }
}

\date{Submitted 2016 February 22}

\pagerange{\pageref{firstpage}--\pageref{lastpage}} \pubyear{2016}

\begin{document}
\label{firstpage}
\maketitle

\begin{abstract}
We present a model for describing the general structure of molecular clouds (MCs) at early evolutionary stages in terms of their mass-size relationship. Sizes are defined through threshold levels at which equipartitions between gravitational, turbulent and thermal energy $|W| \sim f(E_{\rm kin} + E_{\rm th})$  take place, adopting interdependent scaling relations of velocity dispersion and density and assuming a lognormal density distribution at each scale. Variations of the equipartition coefficient $1\le f\le 4$ allow for modelling of star-forming regions at scales within the size range of typical MCs ($\gtrsim4$~pc). Best fits are obtained for regions with low or no star formation (Pipe, Polaris) as well for such with star-forming activity but with nearly lognormal distribution of column density (Rosette). An additional numerical test of the model suggests its applicability to cloud evolutionary times prior to the formation of first stars.
\end{abstract}


\section{Introduction}
Characterizing the general structure of star-forming regions is an issue which is subject of and worth of intensive study. The molecular clouds (MCs) associated with them are the original sites of star formation \citep[for reviews, see][]{MacLow_Klessen_04, McKee_Ostriker_07, Klessen_Glover_14}. Denser fragments of MCs, often labeled `cores' and/or `clumps', turn out to have mass distributions similar or identical to the initial stellar mass function \citep[][Table 1]{ALL_07, VDK_13}. This raises the problem whether there is a link between the general structure of a cloud and its star-forming properties. Some basic indicators of general cloud structure are, for instance: i) the existence of scaling relations of velocity dispersion and density \citep{Larson_81, Solomon_ea_87, Heyer_ea_09, KLN_13}; and, ii) the probability distribution of column density whose shape could be close to lognormal \citep{LAL_11, Brunt_15}, to a power-law function \citep{LAL_15} or a combination of both \citep{Kainulainen_ea_09}. 
The 
analysis of the indicator ii) is considered as a key to understanding the evolutionary status of the cloud and the dominant processes that govern its physics \citep[see][for discussion]{Schneider_ea_13, Schneider_ea_15a}. By use of the probability distribution function (pdf) one can calculate masses within chosen density thresholds and -- defining effective size in some way, -- study the intra-cloud mass-size relationship \citep{LAL_10, BP_ea_12}.

In this Paper we model general MC structure assuming power-law scaling relations of velocity dispersion and density and a lognormal density distribution at each scale. The scales are defined through iso-density contours within which an equipartition between gravitational, kinetic and thermal energy exists. The physical basis and the construction of the model are described in Sect. \ref{Model}. The predicted mass-size relationships and their comparison with observational data for several Galactic star-forming regions are presented in Sect. \ref{Model predictions}. Sect. \ref{Discussion} contains a discussion of the applicability of the model in terms of column-density range, cloud evolutionary stage and size of the star-forming region. A summary of this work is given in Sect. \ref{Summary}.

\section{Model of cloud structure}
\label{Model}
\subsection{Physical framework}
\label{Framework}
The cloud is considered to be at an early evolutionary stage, prior to formation of stars and/or stellar clusters in its densest parts. Its possible age is in the range $t_{1}\gtrsim5$~Myr, corresponding to fully developed supersonic turbulence, and $t_{2}\lesssim18-20$~Myr, corresponding to a global cloud contraction, as suggested by numerical simulations of cloud evolution \citep[e.g.][]{Baner_ea_09, VS_ea_07} or observations of nearby galaxies \citep{Fukui_ea_09, Meidt_ea_15}. Hence the general structure of the cloud is determined primarily by the interaction of supersonic turbulence and gravity. The fully developed turbulence shapes the density and velocity field at any spatial scale $L$ within the inertial range through a cascade possibly driven by the very process of cloud formation \citep{KH_10}. We set conservative limits of the inertial range: $0.1\lesssim L \lesssim 20$~pc. The lower limit is close to the transonic scale and to the typical size of dense (prestellar) cores. The upper limit of 
$20$~pc is adopted to ensure that the gas is mainly molecular and isothermal (with temperature $T=10-20$~K). This estimate is plausible as well if one takes into account that the largest scale of the inertial range is about 3 times less than the injection scale and adopts for the latter $\sim50$~pc, which is above the typical size of giant MCs \citep{Kritsuk_ea_07, Padoan_ea_06}. An equipartition of gravitational vs. kinetic and thermal energy takes place within the mentioned evolutionary stage as gravity slowly takes over toward a global cloud contraction (Fig. 8 in \citealt{VS_ea_07}; see also \citealt{ZA_ea_12}). 

\subsection{Basic assumptions}
\subsubsection{Scaling relations of velocity dispersion and mean density}
\label{Assumptions: scaling relations}
\begin{figure} 
\hspace{3.5em}
\begin{center}  
\includegraphics[width=84mm]{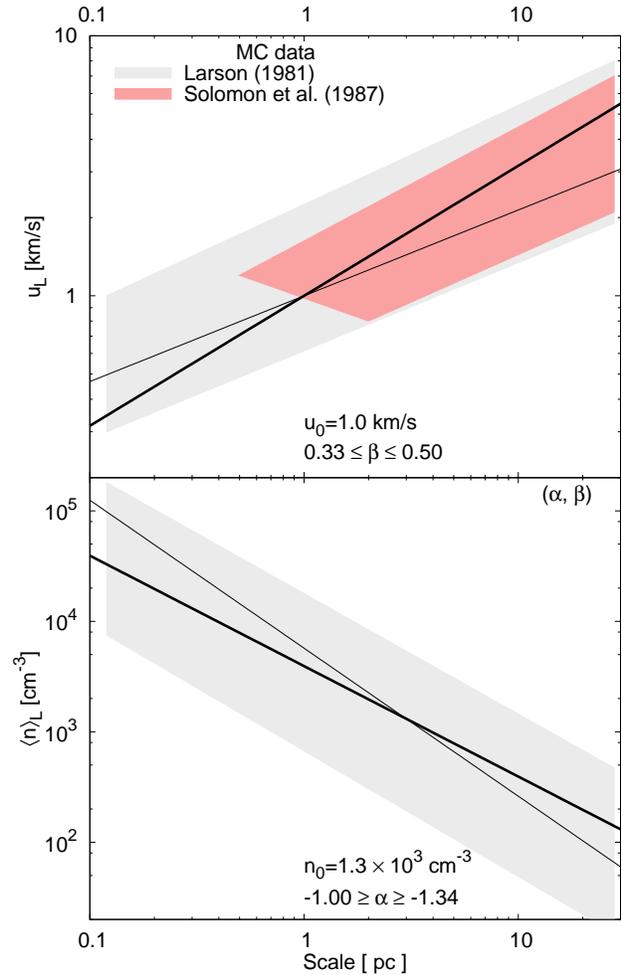}
\vspace{0.9cm}  
\caption{Scaling relations of velocity (top) and mean density (bottom) when the scaling indices $\alpha$ and $\beta$ are varied and for a fixed set of coefficients $(u_0,n_0)$. Shaded areas show the data scatter. The `classical' case $\beta=0.50,~\alpha=-1$ (thick line) and that of shallow velocity scaling $\beta=0.33,~\alpha=-1.34$ (thin line) are plotted.}
\label{fig_scaling_relations}
\end{center} 
\end{figure}

Power-law scaling relations of velocity dispersion $u_L$ and mean density $\langle n \rangle_L$ are assumed to hold within the adopted inertial range. Applied to MCs and cloud fragments, they were initially discovered by \citet{Larson_81} and therefore are often called ``Larson's first and second relations''. In our modelling, we use these relations in the form:
\begin{equation}
\label{eq_Larson_1}
 u_L = u_0\,\Big(\frac{L}{1~{\rm pc}}\Big)^\beta~, 
\end{equation}
\begin{equation}
\label{eq_Larson_2}
 \langle n \rangle_L = n_0\,\Big(\frac{L}{3~{\rm pc}}\Big)^\alpha~. 
\end{equation}
The suggested normalization was chosen in view of the scatter of original data \citep{Larson_81, Solomon_ea_87, Falgarone_McKee_15} and of the possible variations of the scaling index $0.33\lesssim\beta\lesssim0.50$ \citep{Larson_81, HB_04, Padoan_ea_06, Padoan_ea_09} where the classical value of \citet{K41} for incompressible turbulence is taken as a lower limit. The scaling indices $\alpha$ and $\beta$ are interdependent in the proposed model (see our next basic assumption \ref{Assumptions: equipartition}) and thus the variations of $\beta$ generate $-1.3\lesssim\alpha\lesssim-1$. To provide consistency of the mean-density scaling relation with such index values and within the inertial range a higher normalization factor of the scale was adopted in equation \ref{eq_Larson_2} (cf. Fig. \ref{fig_scaling_relations}, bottom). 

Different estimates of the scaling coefficient $u_0$ can be found in the literature \citep{Heyer_ea_09, BP_ea_11a, Shetty_ea_12} while $n_0$ is less studied, in particular, due to the variety of ways to define discrete objects in MCs and their density. Reference values of $u_0$ and $n_0$ yielding scaling relations in agreement with observational data are given in Table \ref{table_var_scaling_coef} and two concrete scaling relations for fixed scaling coefficients are shown in Fig. \ref{fig_scaling_relations}. Note that $u_0= u (L=1~{\rm pc})\gtrsim1$~km/s is typical for dense cloud regions which are possible sites of star formation \citep[see][Fig. 3]{BP_ea_11a}.

\begin{table}
\caption{Fiducial values of the scaling-relations coefficients $u_0$ and $n_0$ allowing variations of indices $\alpha$ and $\beta$ within the confidence ranges.
}
\label{table_var_scaling_coef} 
\begin{center}
\begin{tabular}{cccc}
\hline 
\hline 
$u_0$ & $n_0$ & $-\alpha$ & $\beta$  \\ 
~[ km/s ] & [ \cc ]  & ~ & ~ \\ 
\hline 
1.0&	$1.3\times10^{3}$ &	0.90-1.34&	0.33-0.55 \\
1.0&	$4.4\times10^{3}$ &	0.90-1.34&	0.33-0.55 \\	
1.4&	$1.3\times10^{3}$ &	1.00-1.34&	0.33-0.50 \\
1.4&	$4.4\times10^{3}$ &	0.90-1.34&	0.33-0.55 \\		
\hline 
\hline 
\end{tabular} 
\end{center}
\smallskip 
\end{table}

\subsubsection{Equipartition between gravitational and kinetic energy, including thermal support}
\label{Assumptions: equipartition}
The equipartition relation is described by the equation:
 \begin{equation}
  \label{eq_equipartition}
 |W|=f (E_{\rm kin} + E_{\rm th})~, 
  \end{equation}
where $W$, $E_{\rm kin}$ and $E_{\rm th}$ are gravitational, kinetic (turbulent) and thermal (internal) energy per unit volume $v$ and the coefficient $f$ is taken to vary from unity to 4, i.e. from weakly-gravitating to strongly gravitationally bound entities. As we consider cold molecular gas with $T=10-20$~K, the thermal energy term in the equation above is much less than the gravitational and the kinetic ones and thus contributes only for the fine energy balance.

\citet{BP_06} demonstrated (see Sect. 3.6 there) that in case of equipartition between gravitational and kinetic energy the scaling indices $\alpha$ and $\beta$ are interdependent:
\begin{equation}
 \label{eq_alpha_beta_relation}
 \beta=\frac{\alpha+2}{2}~~.
\end{equation}

Interestingly, \citet{BP_VS_95} found from numerical simulations that equipartitions of this type hold for regions of various size in turbulent interstellar medium, defined by some density threshold.

\subsubsection{Lognormal density distribution at each scale}
A lognormal volumetric distribution of density is found in numerous numerical simulations of supersonic turbulence \citep[e.g.][]{Klessen_00, LKM_03, Kritsuk_ea_07, Federrath_ea_10} and is described through a standard lognormal pdf: 
\begin{equation}
\label{eq_v_pdf}
p_v(s)\,ds=\frac{1}{\sqrt{2\pi \sigma^2}}\,\exp{\Bigg[-\frac{1}{2}\bigg( \frac{s -s_{{\rm max,}\,v}}{\sigma}\bigg)^2 \Bigg]}\,ds~,
\end{equation}
where $s=\ln[n/\langle n \rangle_L]$ is the log density, $s_{\rm max}$ is the distribution peak and $\sigma$ is the standard deviation. The latter two parameters are interdependent (see \citealt{VS_94}) and are determined from the sonic Mach number ${\cal M}=u_L/c_{\rm s}$ ($c_{\rm s}$ is the sound speed) and turbulence forcing parameter $b$: 
\begin{equation}
\label{eq_sigma_PDF}
\sigma^2={\rm ln}\,(1+b^2\,{\cal M}^2)~,~~~s_{{\rm max},\,v}=-\frac{\sigma^2}{2}
\end{equation}

In our model we also use the mass-weighted log-density pdf:
\begin{equation}
\label{eq_m_pdf}
p_m(s)\,ds=\frac{1}{\sqrt{2\pi \sigma^2}}\,\exp{\Bigg[-\frac{1}{2}\bigg( \frac{s -s_{{\rm max,}\,m}}{\sigma}\bigg)^2 \Bigg]}\,ds~,
\end{equation}
where $s_{{\rm max,}\,m}=-s_{{\rm max,}\,v}$ \citep[see][Sect. 3.3.1]{LKM_03}.

The turbulence forcing parameter $b$ is taken to span values between $0.33$, for purely solenoidal forcing, and $0.42$, for a natural mixture between solenoidal and compressive modes \citep{FKS_08, Federrath_ea_10, Konstandin_ea_15}.

\subsubsection{Introduction of physical scale}
\label{Physical scale}
Introducing a characteristic turbulent scale $L$ is straightforward. We define it as the linear size of a cube within which velocity dispersion and mean density are calculated according the assumed scaling relations (equations \ref{eq_Larson_1} and \ref{eq_Larson_2}). This quantity is essentially statistical since it is linked to statistical properties of fully developed turbulence. Our second assumption (Sect. \ref{Assumptions: equipartition}) requires another, deterministic definition of scale, through the total volume of regions wherein the balance of gravitational vs. kinetic and thermal energies is achieved. Inspired by the finding of \citet{BP_VS_95}, we define such {\it physical scale} $L_t\le L$ as the effective size of the sum of all regions delineated by log-density threshold level $t$ at which equation \ref{eq_equipartition} is satisfied. 

\begin{figure} 
\begin{center}
\includegraphics[width=50mm]{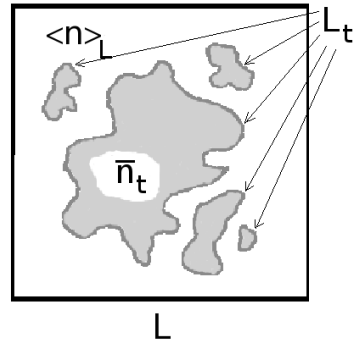}
\vspace{0.2cm}  
\caption{The notion of physical scale $L_t$: the effective size of the sum of all regions (grey `cloudlets') delineated by given log-density threshold level $t$ (thick dark-grey line) and incorporated within turbulent scale $L$.}
\label{fig_physical_scale}
\end{center}
\end{figure}

We stress that the notion of physical scale is not to be confused in any way with a connected region or a clump (Fig. \ref{fig_physical_scale}). To create an intuitive reference to observable objects, we label the regions included in a physical scale `cloudlets'. The size of a single cloudlet can vary from a few pixels on a map (or, in the 3D case, numerical cube) up to a size of whole clouds.

\subsection{Parameters of the cloudlets in equipartition}
Let $t$ be the threshold level at which the equipartition of energies (equation \ref{eq_equipartition}) is achieved. Then the mean log density of set of cloudlets delineated by $t$ is:
\begin{eqnarray}
 \label{eq_mean_log_density_t}
 \overline{s}_t & = & \ln\,[\overline{n}_t/\langle n \rangle_L] =\int\limits_{t}^\infty s\,p_v(s)\,ds \nonumber \\
 ~ & = & \frac{1}{\sqrt{2\pi\sigma^2}} \int\limits_{t}^\infty s \exp{\Bigg[-\frac{1}{2}\bigg( \frac{s + \sigma^2/2}{\sigma}\bigg)^2 \Bigg]}\,ds~.
\end{eqnarray}

The total mass of these cloudlets is defined through the mass of the turbulent scale $M_L=\mu\langle n \rangle_L L^3$ using the mass-weighted log-density pdf (equation \ref{eq_m_pdf}):
\begin{eqnarray}
 \label{eq_mass_t}
 M_t & = & M_L \frac{1}{\sqrt{2\pi\sigma^2}} \int\limits_{t}^\infty \exp{\Bigg[-\frac{1}{2}\bigg( \frac{s - \sigma^2/2}{\sigma}\bigg)^2 \Bigg]}\,ds \nonumber \\
 ~ & = & M_L\, \frac{1}{2}{\rm erfc}(t_m)~,
\end{eqnarray}
where $t_m=(t-\sigma^2/2)/(\sigma \sqrt{2})$ and a mean particle mass $\mu=1.37 m_{\rm u}$ is adopted which accounts for Galactic abundances of atomic and molecular hydrogen and heavier elements \citep{Draine_11}. 

The size of the physical scale is calculated straightforwardly from its volume $V_t$:
\begin{equation}
 \label{eq_size_t}
 L_t = V_t^{1/3}=  L\,\bigg[\int\limits_{t}^\infty p_v(s)\,ds \bigg]^{1/3} = L\,\bigg[ \frac{1}{2}{\rm erfc}(t_v)\bigg]^{1/3}~,
\end{equation} 
where $t_v=(t+\sigma^2/2)/(\sigma \sqrt{2})$.

\subsection{The equipartition equation}
By use of expressions (\ref{eq_mean_log_density_t})-(\ref{eq_size_t}) the gravitational, kinetic and thermal energies per unit volume read:
\begin{equation}
 \label{eq_energy_expressions}
 |W|=\frac{3z_c}{5}G\frac{M_t}{L_t}\mu \overline{n}_t~,~~E_{\rm kin}=\frac{1}{2}\mu \overline{n}_t u_t^2~,~~E_{\rm th}=\frac{3}{2}\overline{n}_t \Re T,
\end{equation}
where $\overline{n}_t=\langle n \rangle_L\,\exp(\overline{s}_t)$ and $u_t=u_L(L_t/L)^\beta$ are the mean density and velocity dispersion of cloudlets, respectively, and $\Re$ is the universal gas constant. The coefficient $1\le z_c\le 2$ accounts for the contribution of the mass outside the cloudlets to their total gravitational energy. In this work, we adopt $z_c=1.5$ like in \citet{DVK_11}. 

Now the equipartition equation (\ref{eq_equipartition}) can be written in terms of the scaling indices $\alpha$ and $\beta$: 
\begin{eqnarray}
  \label{eq_equipartition_equation}
  \noindent\frac{3z_c}{5} G \mu n_0 \bigg(\frac{L}{3~{\rm pc}}\bigg)^\alpha \bigg(\frac{L}{1~{\rm pc}}\bigg)^2 \bigg[0.5^{2/3}\frac{{\rm erfc}(t_m)}{{\rm erfc}(t_v)}\bigg] = \nonumber \\
  ~~~~~~ = \frac{f}{2}\bigg[u_0^2 \bigg(\frac{L}{1~{\rm pc}}\bigg)^{2\beta} \big[ 0.5\, {\rm erfc}(t_v)\big]^{2\beta/3} + 3 \Re T \bigg]~,
\end{eqnarray}
which becomes an equation for $\beta$ through equation (\ref{eq_alpha_beta_relation}). The other free parameters of the model are $f$, $u_0$, $n_0$ and the turbulence forcing parameter $b$ which is implicitly present in the error functions. Varying the threshold level $t$, one can find (if existing) a solution for fixed values of the scaling indices of velocity and mean density. The dynamic range of $b$ turns out to be constrained in the predominantly solenoidal regime -- no solutions were obtained for compressive forcing ($b\ge0.42$).

\section{Model predictions}
\label{Model predictions}
\subsection{Mass-size relationship}
\label{mass-size_relationship}
Mass-size diagrams are often used as a tool to study general structure of MCs and star-forming regions \citep{Lada_ea_08, LAL_10, Kauffmann_ea_10b, Beaumont_ea_12, Shetty_ea_12}. Basically a power-law mass-size relationship $M\propto L^{\gamma}$ has been found, where the index $\gamma$ is constant or changes slowly with the effective size $L$. However, the definition of $L$ introduced by various authors is different. The work of \citet{Kauffmann_ea_10b} makes use of the {\sc Dendrogram} clump-finding algorithm \citep{Rosolowsky_ea_08} and applies it to a set of MC maps, obtained from dust-continuum and dust-emission observations. These authors analyse the mass-size relationship of the extracted objects which build up a hierarchy of embedded connected regions with increasing mean density. \citet{LAL_10} study dust-extinction maps of nearby star-forming regions and delineate structures of given effective size varying stepwise the level of constant absorption. These objects are similar to the physical scales in 
our model (Sect. \ref{Physical scale}) -- in both approaches a fixed density threshold defines a set of cloudlets to which a single effective size is ascribed. However, in contrast to the work of \citet{LAL_10}, the threshold value $t$ here is not arbitrary but is determined by the required equipartition of energies (equation \ref{eq_equipartition}).  

\begin{figure} 
\hspace{3.5em}
\begin{center}
\includegraphics[width=84mm]{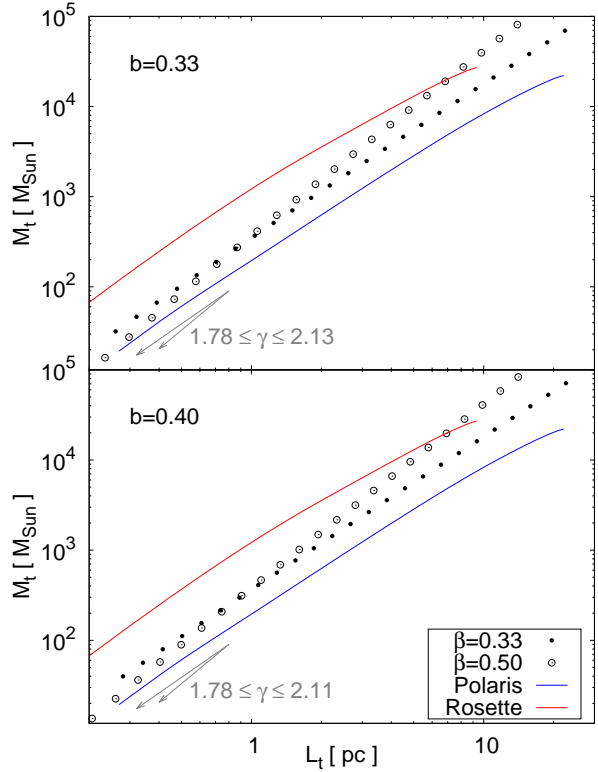}  % width=.8\textwidth
\vspace{0.2cm}  
\caption{Mass-size diagrams of models with $f=2$, $n_0=1.3\times10^3$~\cc, $u_0=1.4$~km/s. The slopes at small scales ($\gamma\sim0.1$~pc), for $\beta=0.33$ and $\beta=0.50$, are indicated with arrows. The mass-size relationships in Polaris (derived from {\it Planck} data) and in Rosette ({\it Herschel} data), are plotted for comparison (see text).}
\label{fig_MR_var_beta}
\end{center}
\end{figure}

The mass-size relationship in our model is defined as a power-law relation $M_t\propto L_t^{\gamma}$ between the physical scale and the mass $M_t$ enclosed therein. From small to large scales within the inertial range $0.1~{\rm pc}\lesssim L_t \lesssim 20$~pc and for any combination of free model parameters, the scaling index $\gamma(0.1)\ge \gamma \ge \gamma(20)$ decreases monotonically  -- see the plots $\gamma(L_t)$ in Appendix \ref{Appendix_gamma_variations}. Yet the variation of $\gamma$ in each considered model case does not exceed 0.3 dex (Table \ref{table_var_gamma}). 

Fig. \ref{fig_MR_var_beta} illustrates the effect of varying the velocity scaling index and the turbulent forcing parameter. The latter evidently does not affect the predicted mass-size relationship for fixed $\beta$ (cf. top and bottom panel). On the other hand, gradual change of $\beta$ from $0.33$ (incompressible turbulence) to $0.50$ leads to corresponding general steepening of the slopes $\gamma$. When mass-size relationships are derived through imposing extinction or column density thresholds, the plausible slopes at small scales are $\le 2$, in view of the properties of the extinction/column density pdf \citep{BP_ea_12}. This is illustrated also by observational mass-size relationships for a region with (Rosette) and without star-forming activity (Polaris) in Fig. \ref{fig_MR_var_beta}. Such slopes can be reproduced by models with $\beta=0.33$ (or a bit larger; Table \ref{table_var_gamma}, Column 5 \& 6) which we take into further consideration. 

\begin{table}
\caption{Variation of the slope $\gamma$ of the mass-size relationship from small ($L_t\sim0.1$~pc) to large scales ($L_t\sim20$~pc).}
\label{table_var_gamma} 
\begin{center}
\begin{tabular}{cc@{~~~}c@{~~~}ccccc}
\hline 
\hline 
~ & ~ & ~ & ~ & \multicolumn{2}{c}{$\beta=0.33$} & \multicolumn{2}{c}{$\beta=0.50$}  \\ 
$f$ & $n_0$ & $u_0$ & $b$ & $\gamma(0.1)$ & $\gamma(20)$ & $\gamma(0.1)$ & $\gamma(20)$\\ 
\hline 
1 &	1.3 &	1.4 &	0.33 &	1.72 &	1.66 &	2.08 &	2.01\\
2 &	1.3 &	1.0 &	0.33 &	1.78 &	1.67 &	2.18 &	2.01\\
2 &	1.3 &	1.4 &	0.33 &	1.80 &	1.68 &	2.21 &	2.00\\
3 &	1.3 &	1.4 &	0.33 &	1.88 &	1.69 &	2.29 &	2.03\\
3 &	4.4 &	1.4 &	0.33 &	1.72 &	1.66 &	2.07 &	2.00\\
4 &	1.3 &	1.0 &	0.33 &	1.93 &	1.68 &	2.34 &	2.04\\
4 &	1.3 &	1.4 &	0.33 &	1.92 &	1.71 &	2.31 &	2.04\\
4 &	4.4 &	1.4 &	0.33 &	1.80 &	1.67 &	2.11 &	~2.01 \vspace{6pt}\\
1 &	1.3 &	1.4 &	0.40 &	1.70 &	1.66 &	2.04 &	2.00\\
2 &	1.3 &	1.0 &	0.40 &	1.73 &	1.66 &	2.11 &	2.01\\
2 &	1.3 &	1.4 &	0.40 &	1.76 &	1.67 &	2.15 &	2.01\\
3 &	1.3 &	1.4 &	0.40 &	1.81 &	1.68 &	2.20 &	2.03\\
3 &	4.4 &	1.4 &	0.40 &	1.69 &	1.66 &	2.02 &	2.00\\
4 &	1.3 &	1.0 &	0.40 &	1.86 &	1.69 &	2.27 &	2.02\\
4 &	1.3 &	1.4 &	0.40 &	1.85 &	1.69 &	2.29 &	2.02\\
4 &	4.4 &	1.4 &	0.40 &	1.74 &	1.66 &	2.09 &	2.01\\
\hline 
\hline 
\end{tabular} 
\end{center}
\smallskip 
\end{table}

\subsection{Comparison with recent observations}
\label{Observational test}
\begin{figure} 
\hspace{3.5em}
\begin{center}
\includegraphics[width=83mm]{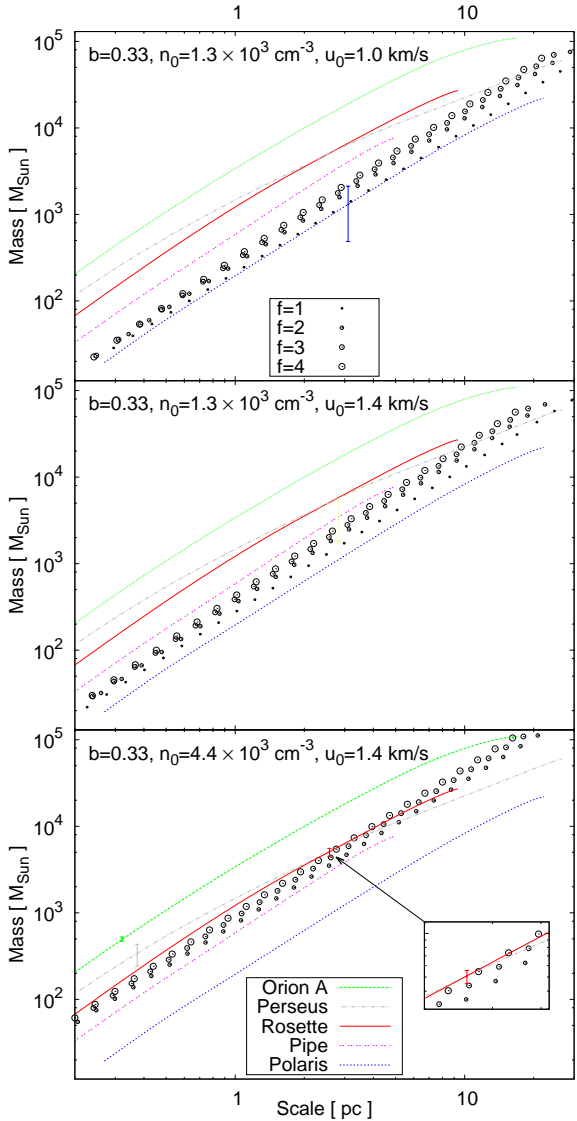}
\vspace{0.7cm}  
\caption{Mass-size ($M_t$ -- $L_t$) relationships from models with $\beta=0.33$ compared with those of several Galactic star-forming regions, derived from {\it Planck} observations. Typical uncertainties of the mass estimates due to uncertainties or gradients of distance to/within given region are shown.}
\label{fig_MR_var_rho0_u0_panel}
\end{center}
\end{figure}

A more detailed observational test of the proposed model is made by use of publicly available {\it Planck} dust-opacity maps\footnote{Based on observations obtained with Planck (http://www.esa.int/Planck), an ESA science mission with instruments and contributions directly funded by ESA Member States, NASA, and Canada.} on the Galactic regions Polaris, Perseus, Pipe and Orion A and of {\it Herschel} data on Rosette \citep{Schneider_ea_12}. Those regions were selected to represent a wide variety of star-forming conditions: a diffuse medium with no signs of star formation (Polaris), a molecular cloud with a few identified young stellar objects (Pipe Nebula), a site of ongoing low- and intermediate-mass star formation (Perseus) and evolved giant MC complexes with star formation (Rosette, Orion A). The mass-size relationships (Fig. \ref{fig_MR_var_rho0_u0_panel}) were obtained from the column-density pdfs ($N$-pdfs) by imposing stepwise thresholds of decreasing column density, like in \citet{LAL_10}. The 
uncertainties of the mass estimates reflect the uncertainties of distance or of distance gradient within the given region \citep[in the case of Perseus, see][]{Schlafly_ea_14}. More information on the selected regions is given in Appendix \ref{Appendix_SFRs} to which we refer the reader.

The ability of our model to describe the general structure of a given star-forming region is quantified through the mass scaling index at small scales $\gamma(0.1)$ and the upper scale $L_{\rm dev}$ of deviation of the model when the observational mass-size relationship is fitted at small scales. The values of $\gamma(0.1)$ (Table \ref{table_var_gamma}, Column 5) are basically consistent with the data in all studied regions. Thus a good agreement could be achieved through variation of the free model parameters (Fig. \ref{fig_MR_var_rho0_u0_panel}) and provided that $L_{\rm dev}$ is at least several pc, i.e. within the size range of typical MCs. 

As mentioned in the previous section, the turbulence forcing parameter does not affect the mass-size relationships and therefore only models with purely solenoidal forcing ($b=0.33$) are shown. Variations of the velocity scaling coefficient $u_0$ toward the higher value $1.4$~km/s (cf. Table \ref{table_var_scaling_coef}) produce small increase of the model masses at a fixed $L_t$ (Fig. \ref{fig_MR_var_rho0_u0_panel}, cf. top and middle panels). Variations of the density scaling coefficient $n_0$ lead to a shift of the mass-size relationships by a factor of 2 or 3 toward higher masses (Fig. \ref{fig_MR_var_rho0_u0_panel}, cf. middle and bottom panels). 

Variations of the equipartition coefficient $f$ evidently allow for modelling of three particular regions at scales $L\le L_{\rm dev}$ where $L_{\rm dev}$ is at least several pc. Models of weakly self-gravitating clouds ($f=1$) and velocity scaling coefficient $u_0=1$~km/s describe very well the structure of internal regions of Polaris Flare (Fig. \ref{fig_MR_var_rho0_u0_panel}, top). That should be expected in view of the sparsity of dense, possibly gravitating cloud cores in this region \citep{Andre_ea_10}. The Pipe Nebula could be described either with strongly gravitationally bound models of lower density scaling coefficient ($f=4$, $n_0=1.3\times10^3$~\cc) or with weakly self-gravitating, but denser models ($f=1$, $n_0=4.4\times10^3$~\cc; Fig. \ref{fig_MR_var_rho0_u0_panel}, middle and bottom panels). This follows from the equipartition equation (\ref{eq_equipartition_equation}) where the increase of $f$ is physically equivalent to the increase of the mean density through $n_0$. The structure of the 
internal parts of the Rosette region could be approximated only by models where the medium is strongly gravitational bound and dense ($f=4,~n_0=4.4\times10^3$~\cc). The latter clearly cannot reproduce the structure of a huge star-forming complex like Orion A, although its index $\gamma(0.1)$ is in the range specified in Table \ref{table_var_gamma}. A possible explanation could be that the whole physical picture in this region is essentially different than the basic assumptions of the model. Orion A is an evolved complex with active but not recent star formation which has been propagating through it within the last dozen of Myr. Numerous dense prestellar cores are detected wherein most of the mass at small scales is concentrated \citep[see][for review, and Sect. 4.2]{Bally_08}. Nevertheless, small scales in this region, with sizes up to several pc, could be speculatively fitted through extreme increasing of the scaling coefficients $n_0$ and $u_0$ beyond the limits which yield scaling relations 
consistent with observational data (see Fig. \ref{fig_scaling_relations}).

In obvious contrast with the structure of other selected regions, Perseus is characterized by a substantial change of the slope of the mass-size relationship from small to large scales. We return to this issue in Sect. \ref{gamma_variations}.

\section{Discussion on the model applicability}
\label{Discussion}

\subsection{The velocity-dispersion scaling index}

A traditional interpretation of the first Larson's relation (equation \ref{eq_Larson_1}) with scaling index $\beta\sim 0.33$ is that the interstellar medium is dominated by subsonic flows while values $\beta\sim0.50$ are held as indicative for highly compressible supersonic turbulence. However, such claims are justified largely by results from numerical simulations of isothermal {\it non self-gravitating} media \citep{Kritsuk_ea_07, Federrath_ea_10}. Also, the two Larson's relations should be considered interdependent even in a purely turbulent case -- through the scaling of the density-weighted velocity dispersion $\rho^{1/3}v$ which is sensitive to the driving mode \citep[solenoidal or compressive; see][for discussion]{Federrath_13}. In the model, proposed here, this interdependence might be additionally modified by gravity, through the assumed equipartitions at each scale in the inertial range. For instance, \citet{Stanchev_ea_15} found $\rho^{1/3}v \propto L^{-2/3}$ in Perseus region, under the 
assumption of equipartition between gravitational and turbulent energy. This is nearly consistent with $\rho^{1/3}v \propto L^{-(2+\beta)/3}$ in this work, given that $\beta=0.33$.

In fact, some numerical works on magnetized clouds yield even shallower velocity power spectrum than predicted in the Kolmogorov theory. For instance, \citet{Collins_ea_12} measured $\beta= 0.23-0.29$ and \citet{Kritsuk_ea_09} obtained $\beta=0.25-0.31$, in consistence with \citet{Lemaster_Stone_09}.

In view of the abovementioned, models with velocity-scaling index $\beta=0.33$ could be considered appropriate to describe general structure of compressible turbulent molecular clouds with an essential role of self-gravity in the energy budget.
 
\subsection{Variety of cloud conditions}

\begin{figure} 
\hspace{3.5em}
\begin{center}
\includegraphics[width=83mm]{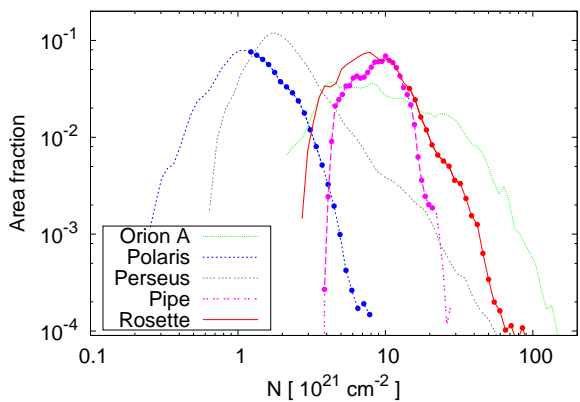}
\vspace{0.7cm}  
\caption{$N$-pdfs of the selected regions, extracted from {\it Planck} and {\it Herschel} (Rosette) data. Those parts that yield mass-size relationships our model is consistent with are shown with bullets.}
\label{fig_SF_regions_Npdfs}
\end{center}
\end{figure}

The obtained values of $L_{\rm dev}$ from fitting of the mass-size relationship in the sampled regions (Table \ref{table_model_agreement}) are within the size range of typical MCs \citep{BT_07}. This suggests that the model is appropriate for description of the dense molecular phase in star-forming complexes. As expected in view of the physical framework of the model (Sect. \ref{Framework}), best fits are obtained for regions with sparse or with no star formation at all: Pipe and Polaris. Their $N$-pdfs have nearly lognormal shapes (Fig. \ref{fig_SF_regions_Npdfs}) with tiny powel-law (PL) tails of very steep slopes, which is typical for inactive complexes \citep{Kainulainen_ea_09, Schneider_ea_15b}. Large scales in Polaris with a mass-size relationship that cannot be fitted ($L>L_{\rm dev}$) correspond to column-density range $N\lesssim 1\times10^{21}$~\sqc~wherein the assumptions for purely molecular phase and, probably, isothermality might be not true \citep{VS_10, Hennebelle_ea_08}. Note, however, that 
the 
lognormal pdfs in our model are defined at abstract scales within the cloud and cannot easily be compared with the {\it single} pdf of the entire cloud which should be considered rather as a superposition of many scale pdfs.

Longer PL tails of the $N$-pdfs with shallower slopes ($\lesssim4$) are indication for gravitational contraction and other processes controlled by gravity which eventually lead to local events of star formation \citep{BP_ea_11b, Schneider_ea_15a}. In our sample, such regions are Rosette, Orion A and Perseus. The PL tail in Rosette is characterized by a slope of $\sim4$ and yields a mass-size relationship that can be modelled up to $L_{\rm dev}\sim4$~pc (Fig. \ref{fig_SF_regions_Npdfs} and Table \ref{table_model_agreement}). In contrary, the model fails to fit the general structure of Perseus and Orion A (Fig. \ref{fig_MR_var_rho0_u0_panel}). Their $N$-pdfs exhibit pronounced PL tails with slopes $2.1$ \citep{Stanchev_ea_15} and $2.7$, correspondingly. Numerical simulations show that such long tails with slopes $\lesssim 3$ characterize strongly self-gravitating media \citep{KNW_11, FK_13, Girichidis_ea_14}. Therefore we revisit the issue of the evolutionary status of the clouds whose structure the model aims 
to represent.   

\begin{table}
\caption{Upper scale $L_{\rm dev}$ of deviation of the model from observational mass-size relationships and its corresponding column density $N(L_{\rm dev})$. The lower limit $N_{\rm obs,\,PL}$ of the power-law tail of the observational $N$-pdf is given in Column 4. The slope of this tail is specified in Column 5.}
\label{table_model_agreement} 
\begin{center}
\begin{tabular}{lcccc}
\hline 
\hline 
Region & $L_{\rm dev}$ & $N(L_{\rm dev})$ & $N_{\rm obs,\,PL}$ & $|$Slope$|$\\ 
~      & [ pc ]     &  [ 10$^{21}$~\sqc ] & [ 10$^{21}$~\sqc ] & \\
\hline 
Polaris	& $\sim15$ &	$~1.2$ &   ~3.7 & $>6$ \\
Pipe	& $\sim~5$ &	$~4.0$ &   13.5 & $>7$ \\
Rosette	& $\sim~4$ &	$14.0$ &   16.0 & $\sim 4$\\
Orion A	&  --       &	--     &   33.0 & 2.7 \\
\hline 
\hline 
\end{tabular} 
\end{center}
\smallskip 
\end{table}

\subsection{Cloud evolutionary stage}
We chose numerical $N$-pdfs from two grid simulations that represent MC evolution at two different evolutionary stages. The simulation S10, analysed in the work of \citet{Shetty_ea_10}, provides a snapshot from the early cloud evolution: well-developed and driven turbulence with increasing contribution of gravity to the energy budget. The simulation S15 from \citet{Stanchev_ea_15} has been aimed to depict the late MC evolution, about and after the formation of first stars. Basic information about the used simulations is given in Table \ref{table_num_simulations}. Their set-ups are comparable in terms of resolution and magnetic field and essentially differ in regard to treatment of turbulence and geometry of clump/cloud formation. For instance, dense regions of gas in S10 form under combined influence of driven random Gaussian velocity field and gravity which is artificially switched on after several dynamical times. On the other hand, cloud formation in the simulation from S15 takes place through collision 
of one-dimensional flows whereas turbulence at the considered late evolutionary stage is due to fluid motions. However, both different treatments of turbulence are realistic from the perspective of evolutionary time. 

\begin{table}
\caption{Summary of the numerical simulations used to test the model. Notation of the reference: S10 - \citet{Shetty_ea_10}; S15 - \citet{Stanchev_ea_15}.}
\label{table_num_simulations} 
\begin{center}
\begin{tabular}{lcc}
\hline 
\hline 
~     &  S10    &  S15 \\
\hline 
Evolutionary time & $0.5$ free-fall times & $\gtrsim15$~Myr \\
Selected area	& $10 \times 10$~pc  &	$40 \times 50$~pc \\
Turbulence & driven & decaying \\
Initial Mach number$^\ast$ & 9.0 & 0.4\\ 
Initial density	& 200~\cc & 1~\cc\\	
Magnetic field & 0.6 $\mu$G & 3 $\mu$G \\
Maximum resolution & $\lesssim 0.01$~pc & $0.03$~pc\\ 
Simulation code	& {\sc ENZO} & {\sc FLASH} \\	
\hline 
\hline 
\end{tabular} 
\end{center}
\smallskip
$\ast$ The initial medium in S10 is isothermal with $T=10$~K while it is warm ($T=5000$~K) in S15.
\end{table}

\begin{figure} 
\hspace{3.5em}
\begin{center}
\includegraphics[width=83mm]{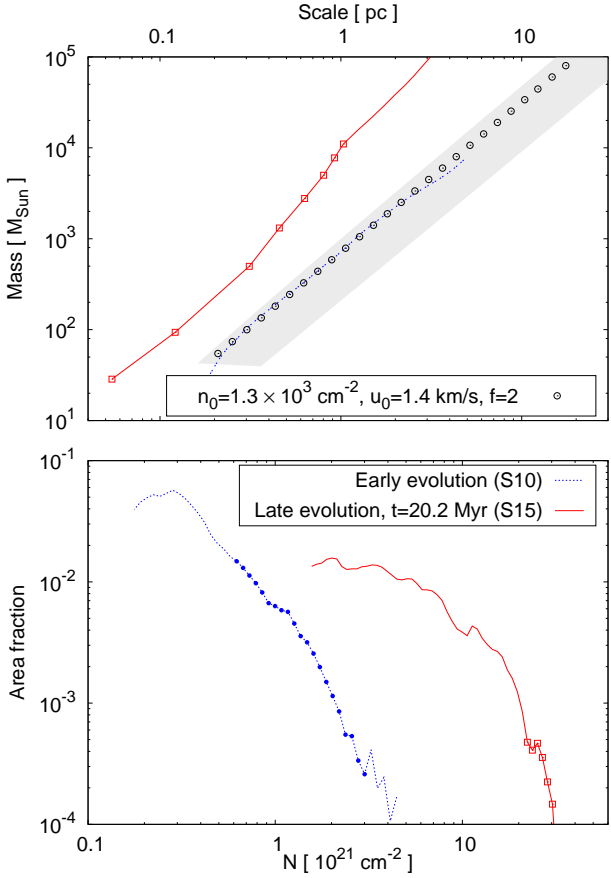}
\vspace{0.7cm}  
\caption{Test of the model from simulations at early (S10) and late (S15) evolutionary stage (see Table \ref{table_num_simulations}). {\it Top:} Mass-size relationships, compared with the predictions of models with $\beta=0.33$, $b=0.33$, $1\le f \le 4$ (shaded area); {\it Bottom:} Numerical $N$-pdfs and the part where the model is consistent with the derived mass-size relationship (bullets). The regime of dense prestellar cores with $n\sim10^4 - 10^5$~\cc~from S15 (squares) is shown in both panels.}
\label{fig_MR_N-pdfs_simulations}
\end{center}
\end{figure}

In Fig. \ref{fig_MR_N-pdfs_simulations} we illustrate an analysis which is analogous to the one performed in the previous Section. As evident from the top panel, the mass-size relationship from S10 is located within the zone, covered by the set of models with $\beta=0.33$, $b=0.33$ and varying $f$, $n_0$ and $u_0$. Best fit is provided by a model with `virial-like' equipartition ($f=2$) as $L_{\rm dev}\sim 3$~pc is about the upper limit of the inertial range in the simulation. The discrepancy at small scales ($\lesssim 0.2$~pc) is probably caused by the end of the inertial range and/or resolution effects -- see the corresponding $N$-pdf tail in the bottom panel. The column-density range wherein the model is consistent with S10 falls entirely in the PL tail with {\it average} slope of about $2$ although the shape is close to part of a lognormal (cf. the $N$-pdf tails in Fig. \ref{fig_SF_regions_Npdfs}). 

On the other hand, the model evidently cannot predict the mass-size relationship from S15 at time $\sim20$~Myr, i.e. after emergence of first stars, even when strong gravitational boundedness ($f=4$) is assumed. We attribute this to the physical conditions in the dense clumps which populate the small scales ($L<1$~pc) in the considered simulation box. Their mean volume densities are at least few times $10^4$~\cc ~(most often, $\sim10^5$~\cc) which hints at their prestellar nature. Typical linewidths of such objects are trans/subsonic and the analysis of their density profiles possibly suggests a lack of equilibrium \citep[for a discussion, see][]{BT_07}. Therefore their physics is inconsistent with the adopted assumptions for supersonic turbulence and energy equipartition. The applicability of our model is thus constrained to the early evolutionary stage of MCs -- with an upper age limit $t\lesssim15$~Myr, about the formation of first stars.

\subsection{Variations of the intracloud mass scaling index}
\label{gamma_variations}
The variation of $\gamma$ in a given model with fixed $f$ is restricted (Table \ref{table_var_gamma}). Thus the discrepancy at $L>L_{\rm dev}$ is caused by a significant drop of the mass-scaling index at larger scales in real star-forming regions. The latter phenomenon could be explained with the characteristics of the cloud's $N$-pdf: lognormal shape with/without PL tail, width of the lognormal part, slope of the PL tail and typical density of transition between both regimes. From Fig. 11 in  \citet{BP_ea_12} one could see that significant variations of $\gamma$ within a region are produced by $N$-pdfs which are combinations of a broad (lognormal) component and a shallow PL tail -- like in Rosette, Perseus and Orion A (Fig. \ref{fig_SF_regions_Npdfs}). The parameters of these two components reflect the balance between turbulence and gravity at different scales in the cloud.

It is physically consistent to expect that this balance is described at different scales by a different type of equipartition (if such is present at all). For example, whereas the densest cores evolve faster and local collapses take place, the global contraction of the cloud starts at a time when first stars have been already formed \citep{VS_ea_07, BP_ea_11b}. In view of this, small scales are to be described by an equipartition with stronger contribution of gravity ($f\ge2$) while the large ones, comparable to the size of entire cloud, should be characterized by $1\le f \le 2$ or less. A combination of models with different choice of $f$ for scales over $L_{\rm dev}$ could reproduce, in principle, the total observational mass-size relationships in some regions. That is evident from an eye inspection of Fig. \ref{fig_MR_var_rho0_u0_panel}. Fixing the other free model parameters, a decrease of $f$ produces less mass at a given scale. In case an observational mass-size relationship is well fitted through a 
model with $f=f^\prime$ at scales $L\le L_{\rm dev}$, it can be successfully reproduced also at $L> L_{\rm dev}$ by a series of models with decreasing $f<f^\prime$.

\section{Summary}
\label{Summary}
We present a model of the general structure of molecular clouds (MCs) at their early evolutionary stage ($5 \lesssim t \lesssim 15$~Myr), characterized by developed supersonic isothermal turbulence and essential contribution of gravity to the energy balance at different spatial scales $L$. Here we consider the range $0.1~{\rm pc}\le L \le 20$~pc, adopting a turbulent injection scale above the typical size of giant MCs. Our model is very sensitive to the evolutionary stage of the cloud, as well as to the properties of its internal turbulence. In particular, it depends on the assumed power-law scaling relations of the velocity dispersion and the mean density, on the equipartition between gravitational and kinetic energy, including thermal support, i.e. $|W| \sim f(E_{\rm kin} + E_{\rm th})$ ($1\le f \le 4$), and on the validity of a lognormal probability density function (pdf) at each turbulent scale $L$. A physical scale $L_t\le L$ is defined as the effective size of the sum of all regions above a log-density 
threshold level $t$ at which the equipartition equation is satisfied. Free parameters of the presented model are the velocity scaling index $\beta$, the coefficients in the scaling relations of velocity ($u_0$) and density ($n_0$), the coefficient of equipartition $f$ and the turbulence forcing parameter $b$. The predictive power of the model is put to test by comparison of the mass-size relationships $L_t - M_t$ with ones, derived from observational column-density pdfs in several Galactic regions of varying star-forming activity as well from two simulations of evolved MCs. 

The results of this study are as follows:
\begin{itemize}
 \item The model predictions of mass-size relationships are not sensitive to the value of $b$, given that turbulence forcing is predominantly compressive, while the variations of the parameters $u_0$ and $n_0$ lead to variations of model masses at a fixed scale $L_t$ within a factor of 3. A velocity scaling index $\beta$ which is significantly larger than the value in Kolmogorov theory ($0.33$) produces mass-size relationships that cannot fit the observational ones. However, it should be not considered as an indication for subsonic turbulence but is rather determined by interplay between gravity and highly compressible turbulence. 
 \item Variations of the equipartition coefficient $1\le f\le 4$ essentially shift the mass range and allow for modelling of some star-forming regions at scales within the size range of typical MCs ($\gtrsim4$~pc). Observed mass-size relationships at larger scales could be reproduced as well by a series of models with decreasing $f$ which is justified in view of the physical state of evolving MCs as revealed from numerical simulations.
 \item The model is able to describe the general structure of regions with low or no star-forming activity, characterized by nearly lognormal $N$-pdf (Polaris, Pipe) as well the structure of some star-forming regions, given that their $N$-pdf is forming a short and steep power-law tail (Rosette). 
 \item Comparisons with two numerical simulations of cloud evolution at different stages show that the model is able to describe the general properties of a medium with driven turbulence and strong self-gravity in the energy balance but prior to eventual star formation -- which is consistent with the basic assumptions.
 \end{itemize}

\vspace{12pt}
{\it Acknowledgement:} T.V. acknowledges support by the {\em Deutsche Forschungsgemeinschaft} (DFG) under grant KL 1358/20-1. We thank R. Shetty and B. K\"{o}rtgen for providing data from their simulations \citep[][respectively]{Shetty_ea_12, Stanchev_ea_15} and O. Stanchev for his support on software and technical issues. \\

% \newpage

\label{lastpage}
\appendix
\section{Variations of $\gamma$ with the physical scale}
\label{Appendix_gamma_variations}
The scaling index $\gamma$ of the mass-size relationship at a considered physical scale $L_t$ is calculated from a linear fit using three points $L_{t, 1}$, $L_{t, 2}$ and $L_{t, 3}$ within a narrow range and $L_t=\sqrt{L_{t, 1}L_{t, 3}}$. The plots for different combinations of free parameters exhibit a monotonic decrease of $\gamma$ from the lower ($0.1$~pc) to the upper ($20$~pc) limit of the inertial range - see Figs. \ref{fig_appendix_gamma_variation_a} and \ref{fig_appendix_gamma_variation_b}.

\begin{figure*} 
\begin{center}
\includegraphics[width=1.\textwidth]{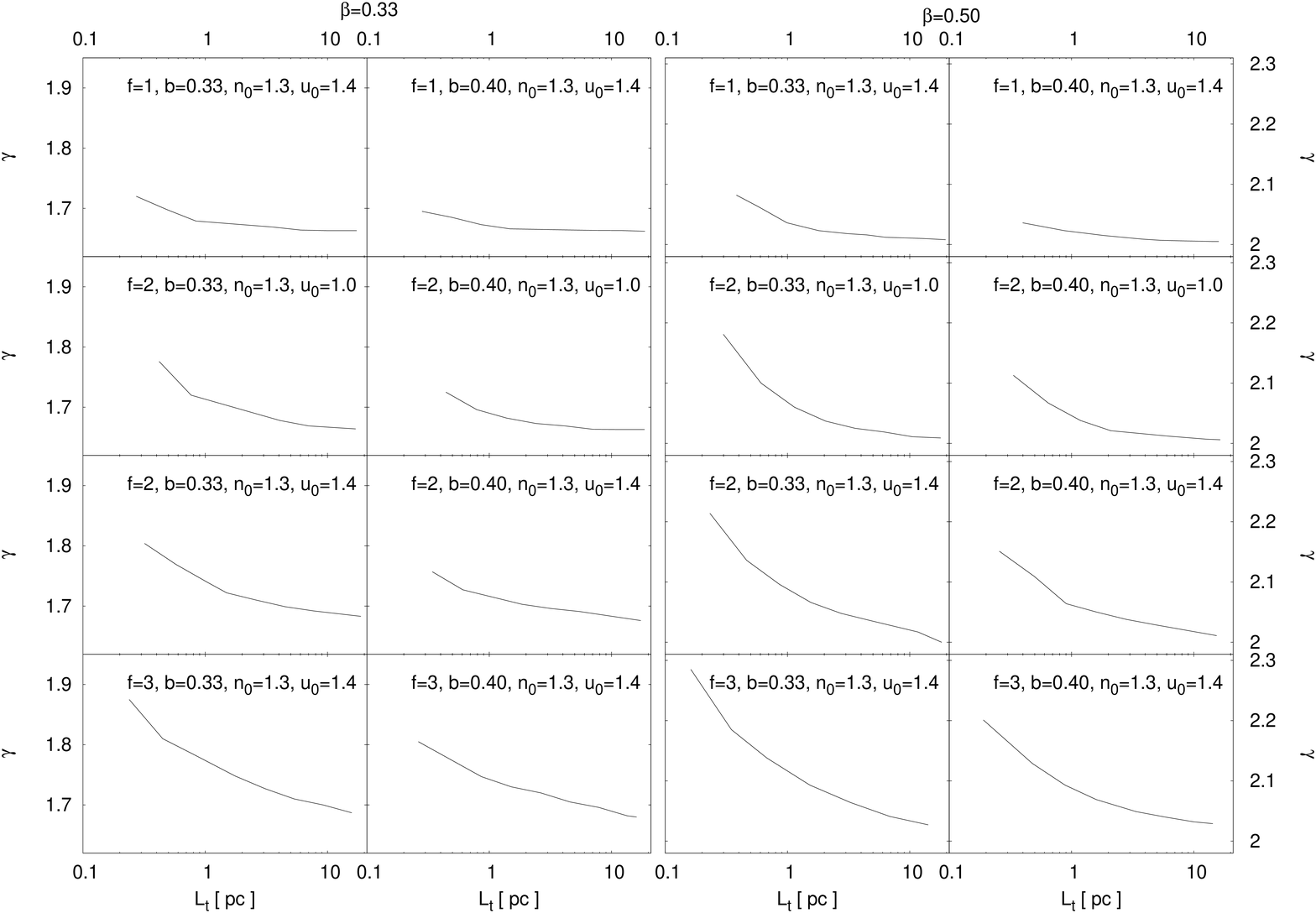}
\vspace{0.4cm}  
\caption{Variations of the scaling index $\gamma$ for models with $\beta=0.33$ (left) and $\beta=0.50$ (right). The sets of free parameter are specified in the upper right corners.}
\label{fig_appendix_gamma_variation_a}
\end{center}
\end{figure*}

\begin{figure*} 
\begin{center}
\includegraphics[width=1.\textwidth]{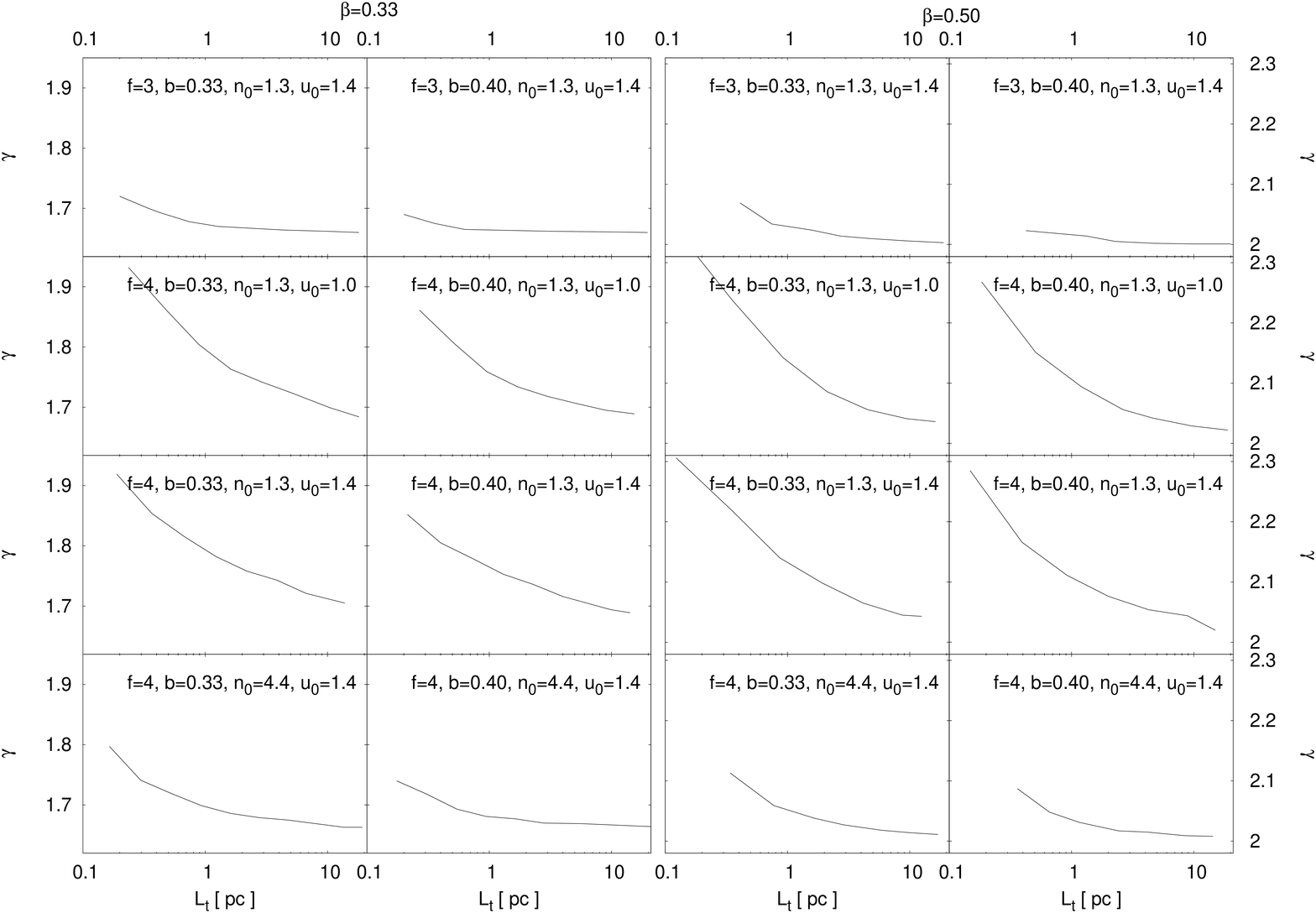}
\vspace{0.4cm}  
\caption{Continuation of Fig. \ref{fig_appendix_gamma_variation_a}.}
\label{fig_appendix_gamma_variation_b}
\end{center}
\end{figure*}

\newpage
\section{Selected zones and mass estimation therein}
\label{Appendix_SFRs}
\begin{figure*} 
\begin{center}
\includegraphics[width=.9\textwidth]{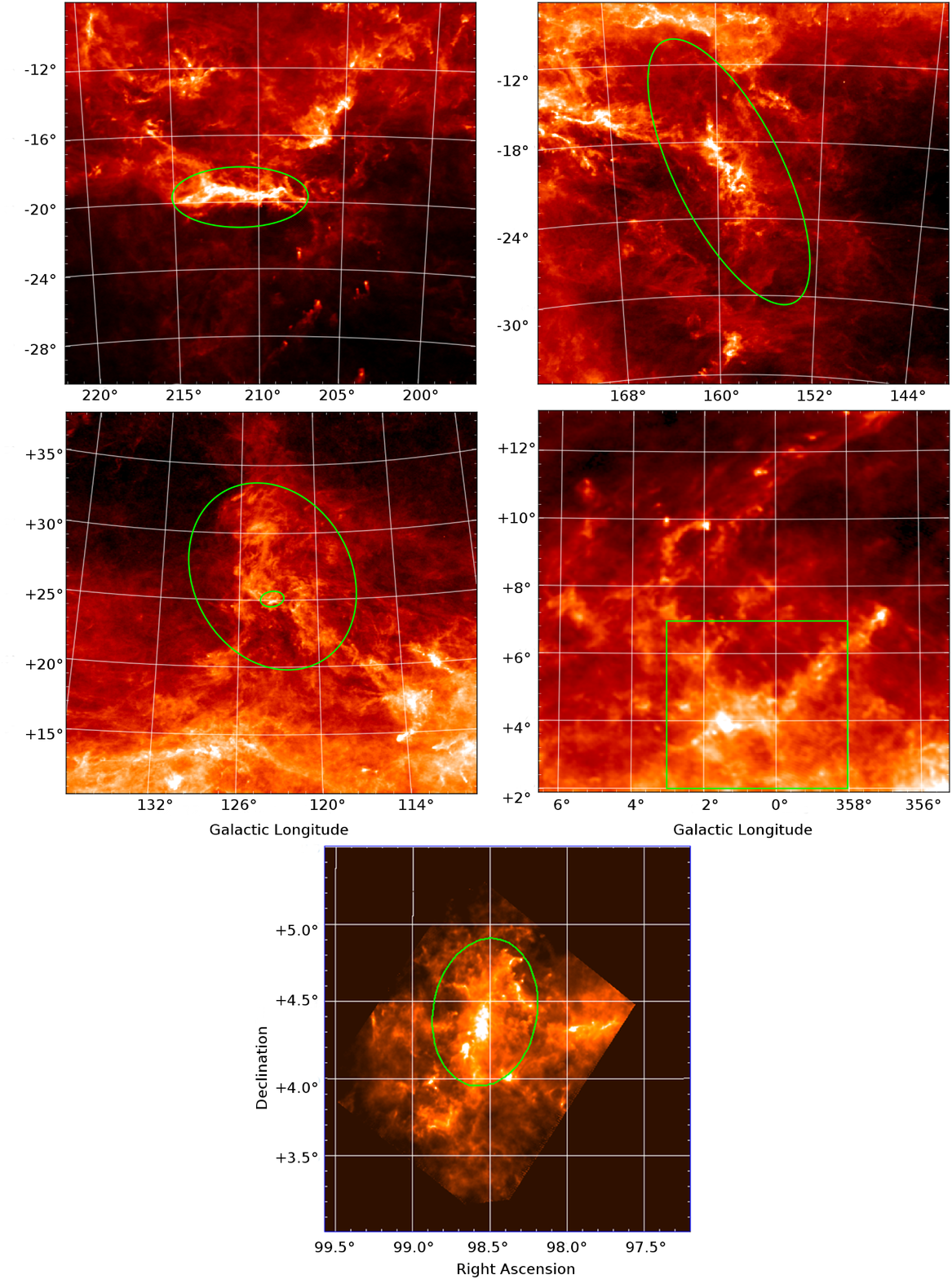}
\vspace{0.4cm}  
\caption{Maps of the selected star-forming regions. The zones used to derive the $N$-pdf are drawn with solid green line. }
\label{fig_SFR_regions}
\end{center}
\end{figure*}

The $N$-pdfs are derived in zones which include MCs associated with the considered star-forming regions (Fig. \ref{fig_SFR_regions}). The zone parameters are specified in Table \ref{table_SFRs_parameters}. The chosen effective sizes and orientations should minimize the effect of fore- or background structures. In case the cloud is apparently and approximately axisymmetric (Orion A, Perseus, Rosette), the major axis was oriented along the axis of symmetry. 

In all but one case the zones were selected from {\it Planck} maps. To estimate masses above given cut-off level of the $N$-pdf, we adopted a linear conversion formula from dust opacity $\tau_{353}$ at 353 GHz to hydrogen column density:
\begin{equation}
 \label{eq_tau_to_N}
 N({\rm H})=C_1 \tau_{353} + C_0~,
\end{equation}
as suggested in \citet{Planck_11}, with coefficients $C_0$ and $C_1$ obtained like in the work of \citet{Stanchev_ea_15}. The possible uncertainty of the calculated column-density is about a factor of 2. Comparing the derived mass-size relationships with our models, we prefer a conservative approach to neglect this uncertainty and cling only to estimates due to uncertainty of distance to the considered star-forming regions (see Table \ref{table_SFRs_parameters}, column 2).      

The mass estimates in Rosette were derived from a column-density map, based on {\it Herschel} observations at four wavelengths (160, 250, 350, and 500 $\mu$m) and constructed as described in \citet{Schneider_ea_12}. 

\begin{table*}
\caption{Parameters of the zones in star-forming regions, selected to test the proposed model. Notation: D = Distance to the region, Ref = reference to the distance estimate, Coor. = Coordinates of the center of the zone, $L_{\rm eff}$ = effective size, $a$ = major semi-axis or square side, PA = position angle of the semi-axis or square side.}
\label{table_SFRs_parameters} 
\begin{center}
\begin{tabular}{lc@{~~~}crrccr}
\hline 
\hline 
Region & Distance & Ref & \multicolumn{2}{c}{Coor.} & $L_{\rm eff}$ & $a$ & PA~~ \\ 
~ & [ pc ] & ~  & $l~~~~$ & $b~~~~$ & [ deg ] & [ deg ] & [ deg ] \\ 
\hline 
Orion A & ~$371\pm {\scriptstyle 10}$ & 1 & 211.104 & $-19.702$ & 3.8 & ~2.6 & ~~0.0 \\
Perseus & ~$260 \pm {\scriptstyle 40}$ & 2 & 159.120 & $-20.340$ & 6.7 & 11.3 & ~24.0 \\
Polaris & ~$150 \pm {\scriptstyle 50}$  & 3 & 123.447 & $26.796$ & 6.3 & ~7.2 & 292.5 \\
Pipe    & ~$130 \pm {\scriptstyle 20}$ & 4 & ~~0.522 & $4.499$ &  -- & ~4.9 & 0.0 \\		
\hline
~ & ~ & ~  & $\alpha\,(2000)$ & $\delta\,(2000)$ & ~ & ~ & ~ \\ 
\hline
Rosette & $1330 \pm {\scriptstyle 50}$ & 5 & 207.015 & $-1.822$ & 0.4 & 0.5 & ~80.6 \\		
\hline 
\hline 
\end{tabular} 
\end{center} 
[1, 5] \citet{LAL_11}; [2] see \citet{Stanchev_ea_15} and references therein; [3] \citet{Bensch_ea_03}; [4] \citet{LAL_06}\\
\end{table*}

\end{document}